# Marco: Configurable Graph-Based Task Solving and Multi-AI Agents Framework for Hardware Design

Chia-Tung Ho, Jing Gong, Yunsheng Bai, Chenhui Deng, Haoxing Ren, Brucek Khailany

NVIDIA, Santa Clara, CA

## Abstract

Hardware design presents numerous challenges stemming from its complexity and advancing technologies. These challenges result in longer turn-around-time (TAT) for optimizing performance, power, area, and cost (PPAC) during synthesis, verification, physical design, and reliability loops. Large Language Models (LLMs) have shown remarkable capacity to comprehend and generate natural language at a massive scale, leading to many potential applications and benefits across various domains. Successful LLM-based agents for hardware design can drastically reduce TAT, leading to faster product cycles, lower costs, improved design reliability and reduced risk of costly errors. In this work, we propose a unified framework, *Marco*, that integrates configurable graph-based task solving with multi-modality and multi-AI agents for chip design by leveraging the natural language and reasoning abilities with collaborative toolkits. Lastly, we demonstrate promising performance, productivity, and efficiency of LLM agents by leveraging the Marco framework on layout optimization, Verilog/design rule checker (DRC) coding, and timing analysis tasks.

## Introduction

The exponential growth in the number of transistors within System-on-Chip (SoC) designs has significantly increased the complexity of meeting Power, Performance, and Area (PPA) requirements in Very-Large-Scale Integration (VLSI). The rising complexity and associated costs underline the necessity to rethink and innovate the EDA processes to accommodate the demands of modern chip architectures. Autonomous agents have long been a research focus in academic and industrial communities across various fields. Recently, AI agents empowered by LLMs [1]-[3] have shown impressive performance in software engineering for solving real world challenging benchmarks (i.e., SWE-Bench, HumanEval) through planning, memory management, actions involving external environment tools. In addition to single-AI agents, many researchers are starting to explore the capabilities of multi-AI agents to optimize and verify complex tasks through collaborative discussions [4], [5]. However, these agent frameworks cannot be directly used for designing hardware because solving hardware tasks requires integrated domain knowledge and specialized hardware design tools to analyze signals, trace signal transitions, and decompose tasks into manageable sub-tasks from circuit architecture and signal transaction perspectives.

The remaining sections are organized as follows. Firstly, we introduce the proposed configurable graph-based task solving and multi-AI agent framework, where agent works on various types of hardware design tasks. Lastly, we present the performances of the developed agents that are supported or enabled by the proposed *Macro* framework in experimental results and conclude our work.

## *Marco*: Configurable Graph-Based Task Solving and Multi-AI Agents Framework

The proposed *Marco* framework, which encompasses graph-based task solving, agent configurations for sub-tasks, and skill/tool configurations for each agent. Fig. 1 illustrates the dynamic and static configurable graph-based task solving, which is flexibly integrated with hardware design knowledge (e.g., circuits, timing, etc.). In the task graph, each node represents a sub-task, and each edge represents the execution or knowledge relationship between nodes. For solving each sub-task, we leverage Autogen [4] to configure single-AI or multi-AI agent with knowledge database, tools, and memory as shown in Fig. 2(a), and Fig. 2(b). Table 1 summarizes the task graph, agent, and LLM configurations of *Marco* framework for various agents. For specification-to-RTL task, VerilogCoder [6] leverages a dynamic task graph for the proposed novel Task and Circuit Relation Graph-based task planner to create a high-quality plan with step-by-step sub-tasks and related circuit information (i.e., signal, signal transition, and single examples). The novel Abstract Syntax Tree-based waveform tracing tool is developed to assist the LLM agent in fixing functional correctness autonomously with the configuration in Fig. 2(b). MCMM timing analysis agent utilizes dynamic task graph to create a task flow that analyzing the timing report for each corner and modes, and then extract key takeaways across MCMM timing reports. The timing path debug agent finds the problematic net, wire, and constraints through static timing debugging task graph as shown in Fig. 1. For RTLFixer [7], cell layout optimizer [8], and DRC Coder [9], we employ single-AI or multi-AI agent configurations in Fig. 2(a) with customized tools, memory, and domain knowledge (Fig. 2(b)).

## Experimental Results

We demonstrate the performance of agents that are supported or enabled by *Marco* framework on optimization, code syntax fixing and generation, and summary and anomaly identification task categories in Table 1.

*Optimization:* The cell layout optimizer [8] generates high-quality cluster constraints to optimize the cell layout PPA and routability with the guidance of designers' expertise and customized tools. The agent not only achieves up to 19.4% smaller cell area, but also generates 23.5% more LVS/DRC clean cell layouts than previous work [10] on a set of sequential cells in industrial $2nm$ technology node.

*Code syntax fixing and generation:* The RTLFixer [7], and VerilogCoder [6] achieve 99% syntax pass-rate, and 94.2% Functional pass-rate on VerilogEval dataset [11] as shown in Fig. 3(a), and Fig. 3(b), respectively. Moreover, Fig. 4 shows that DRC Coder [9] achieves perfect F1 score for a set of industrial DRC rules for standard cell layouts in the advanced technology node. The generated DRC code is evaluated on hundreds of layouts with various types of DRC violations.

*Summary and Anomaly Identification:* The MCMM timing analysis agent achieves an average score of 8.33 out of 10, based on evaluations by experienced engineers on a set of industrial cases, and delivers approximately 60X speedups compared to human engineers as shown in Fig. 5. In Table II, the timing path debug agent resolves 86% of path-level debugging tasks, whereas standard task solving approach fails to resolve any of the tasks.

## Conclusion

The proposed *Marco* framework enables more flexible and domain-specialized methods for hardware design tasks. By leveraging task graph and flexible single-AI/multi-AI agent configurations with domain-specific tools and knowledge, we developed various agents for tasks such as cell layout optimization, Verilog syntax error fixing, Verilog and DRC code generation, and timing debugging on problematic blocks, nets, and wires. The experimental results show impressive performance and efficiency benefits on utilizing collaborative LLM-based agents for hardware design.

Future research directions include: (1) training LLMs with high-quality hardware design data, (2) integrating PPA optimization loops with agentic methodologies, and (3) developing efficient self-learning techniques and memory systems for complex real-world hardware tasks.

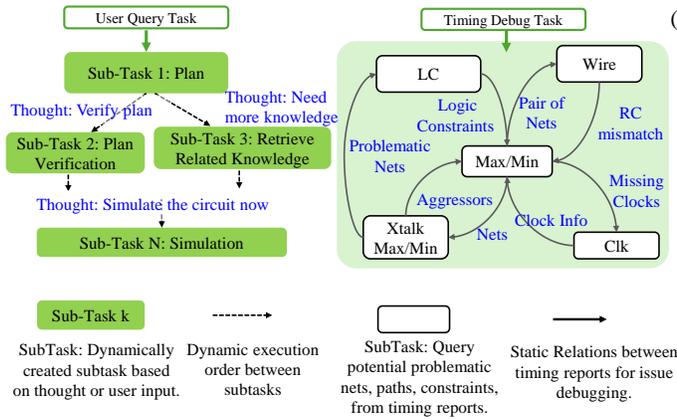
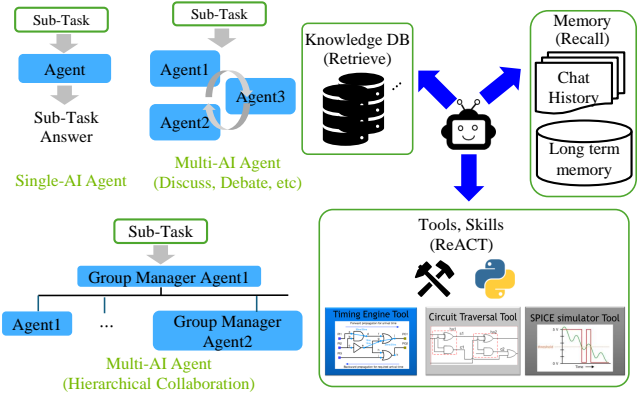

Fig. 1: Graph-Based Task Solving illustration in *Marco*: Configurable Graph-Based Task Solving and Multi-AI Agents Framework.

Fig. 2: (a) Single-AI/Multi-AI Configurations of Sub-Task Node. (b) Agent memory, knowledge database, and tool configurations.

Table I: Task graph, agent configuration, customized tool of *Marco* Framework for various agent implementations for hardware design tasks. In the last column, X (Supported)=The agent can be configured and implemented using *Marco* Framework. V=The agent is implemented on *Marco* Framework.

| Agent Works | Task Category | Configuration of *Marco* Framework | | | Original Agent Work Implementation using *Marco* Framework |
|---|---|---|---|---|---|
| | | Task Graph | Sub-Task Agent Config. | Customized Tools | |
| RTLFixer [7] | Code Syntax Fixing | N/A | Single-AI | RTL Syntax Error RAG Database | ✗ (Supported) |
| Standard Cell Layout Opt. [8] | Optimization | N/A | Single-AI | Cluster Evaluator, Netlist Traverse Tool | ✗ (Supported) |
| MCMM Timing Analysis (Partition/Block-Level) | Summary & Anomaly Identification | Dynamic | Multi-AI | Timing Distribution Calculator, Timing Metric Comparator | ✓ |
| DRC Coder [9] | Code Generation | N/A | Multi-Modality & Multi-AI | Foundry Rule Analysis, Layout DRV Analysis, DRC Code Evaluation | ✓ |
| Timing Path Debug (Path-Level) | Summary & Anomaly Identification | Static (Fig. 1 (a) Right) | Hierarchical Multi-AI | Agentic Timing Report Retrieval | ✓ |
| VerilogCoder [6] | Code Generation | Dynamic | Multi-AI | TCRG Retrieval Tool, AST-Based Waveform Tracing Tool | ✓ |

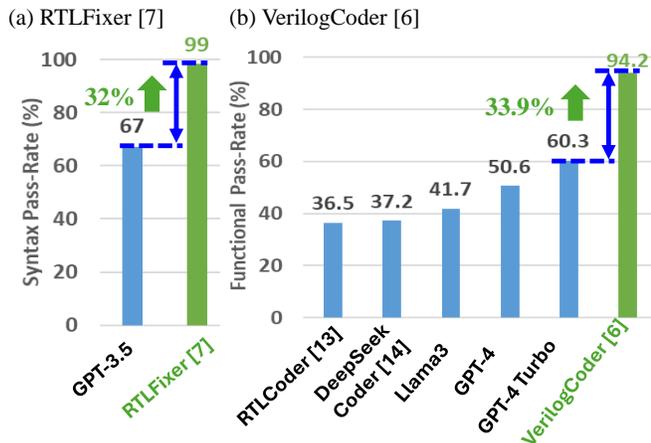
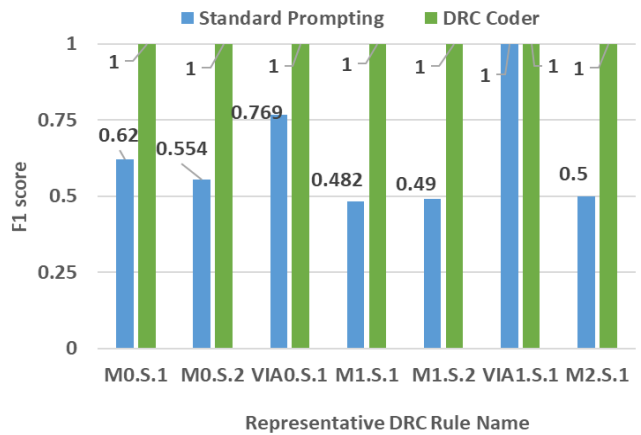

Fig. 3. Verilog syntax fixing and generation performance. (a) Syntax pass-rate on VerilogEval-syntax [12] dataset. The LLM used in RTLFixer [7] agent is GPT-3.5. (b) Functional pass-rates of recent LLMs and the proposed VerilogCoder [6] on VerilogEval-Human v2 [11] dataset. The LLM used in VerilogCoder agent is GPT-4 Turbo.

Fig. 4. DRC Coder [9]: Performance evaluation of DRC code generation using standard prompting and DRC-Coder with GPT-4o across seven design rules. We use F1 Score to evaluate the generated DRC code on DRV detection in an industrial advanced technology node.

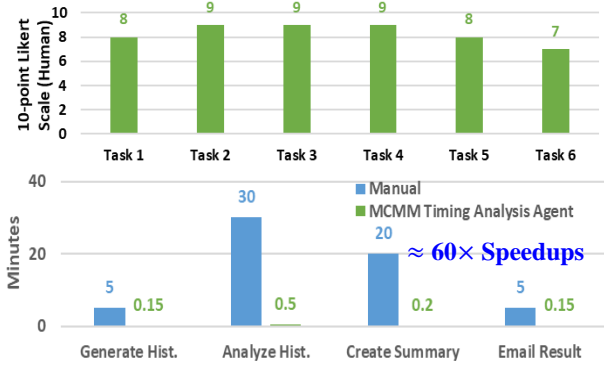

Fig. 5. MCMM Timing Analysis Agent: (Top) Task correctness evaluation of 10-point Likert Scale from human evaluation on 6 timing histogram cases. Here, we omit the details of the 6 tasks due to the tasks are industrial design. (Bottom) MCMM Timing Analysis Agent achieves 60× speedups than experienced human engineer.

Table II: Pass-rate (%) of Timing Path Debug Agent with static task graph solving, and a naïve standard task solving without task graph information. X=failed to solve the task. V=solve the task.

| Task ID | Multi Report Task Description | Required Analyzed Sub-Tasks | Standard Task Solving | Timing Path Debug Agent |
|---|---|---|---|---|
| M1 | Find missing clk signals that have no rise/fall information | max, clk | X | V |
| M2 | Identify pairs of nets with high RC mismatch | max, wire | X | V |
| M3 | Detect unusual constraints between victim and its aggressors | max, xtalk, LC | X | V |
| M4 | Identify unusual RC values between victim and its aggressors | max, wire, xtalk, LC | X | V |
| M5 | Find the constraints of slowest stages with highest RC values | max, wire, xtalk, LC | X | V |
| M6 | Compare each timing table for number of stages, point values and timing mismatch | max | X | X |
| M7 | Task M2 and Task M3 for specific stages in list of paths | max, wire, xtalk, LC | X | V |
| **Avg Pass-rate** | | | **0%** | **86%** |